\documentclass[showpacs,amsmath,amssymb,10pt]{article}
\usepackage{graphicx,color}
\usepackage{amsmath}
\usepackage{amssymb}
\usepackage{amsmath,amsfonts,latexsym,graphicx,amssymb}
\usepackage{amsfonts}
\usepackage{amssymb}
\usepackage{mathrsfs}
\usepackage{amsthm}
\usepackage[utf8]{inputenc}
\usepackage{graphicx}
\usepackage{amsmath}
\newcommand{{\Cd}}{{\mathbb{C}^d}}
\newcommand{{\Rn}}{{\mathbb{R}^n}}

\def\<{\langle}
\def\>{\rangle}
\newtheorem{Theorem}{Theorem}
\newtheorem{cor}{Corollary}

\newtheorem{ex}{Example}

\newtheorem{pro}{Proposition}
\newtheorem{Remark}{Remark}
\date{}

\begin{document}

\title{\textbf{
On Kossakowski construction of positive maps in matrix algebras }}

\author{Dariusz Chru\'sci\'nski\\
Institute of Physics, Faculty of Physics, Astronomy and Informatics \\ Nicolaus Copernicus University, Grudziadzka 5, 87--100 Toru\'n, Poland}

\maketitle

\begin{abstract}
We provide a further analysis of the class of positive maps proposed ten years ago by Kossakowski. In particular we propose a new parametrization which reveals an elegant geometric structure and an interesting interplay between group theory and a certain class of positive maps.
\end{abstract}


\begin{center}
{  \bf Dedicated to Andrzej Kossakowski on his 75th birthday}
\end{center}

\section{Introduction -- a diagonal type positive maps }

Ten year ago in a remarkable paper \cite{Kossak-kule} Kossakowski provided a construction of a family of positive maps in  matrix algebras $M_n(\mathbb{C})$. This construction reproduces many examples of positive maps already known in the literature.
The maps from \cite{Kossak-kule} belong to the following class: let $\{e_0,\ldots,e_{n-1}\}$ denotes an orthonormal basis in $\mathbb{C}^n$ and let $E_{ij} := |e_i\>\<e_j|$. Consider the linear map $\Lambda : M_n(\mathbb{C}) \rightarrow M_n(\mathbb{C})$ defined as follows
\begin{eqnarray}  \label{Lambda}
    \Lambda(E_{ii}) = \sum_{j=0}^{n-1} a_{ij} E_{jj}\ , \ \ \ \
    \Lambda(E_{ij}) = - E_{ij} \ , \ \ \ i \neq j\ .
\end{eqnarray}
where $a_{ij}$ provides a set of complex parameters. In what follows we call the above maps {\em diagonal type maps}, since only diagonal elements $E_{ii}$ are transformed in a non-trivial way.
A map $\Lambda$ is Hermitian, i.e. $[\Lambda(X)]^\dagger = \Lambda(X^\dagger)$ iff $a_{ij} \in \mathbb{R}$. The basic question one poses is:

\begin{center}
 {\em what are conditions for $a_{ij}$ which guarantee that $\Lambda$ is a positive map}.
\end{center}
It is clear that a necessary condition is that all matrix elements $a_{ij} \geq 0$. Observe, that $n\times n$ matrix $A=[a_{ij}]$ with matrix elements $[a_{ij}] \geq 0$ may be considered as a ``classical'' positive linear map $A : \mathbb{R}^n \rightarrow \mathbb{R}^n$. Therefore, formula (\ref{Lambda}) provides a construction of a ``quantum" positive map $\Lambda$ out of the ``classical" map $A$ if ``classical"  conditions $a_{ij}\geq 0$ are completed by a set of suitable ``quantum" conditions.  This problem is easily solvable for $n=2$.  One proves the following

\begin{pro} If $n=2$, then $\Lambda$ is positive if and only if $a_{ij} \geq 0$ and
\begin{equation}\label{Q2}
  \sqrt{a_{00}a_{11} } + \sqrt {a_{01} a_{10}} \geq 1\ .
\end{equation}
Moreover, $\Lambda$ is completely positive if and only if $a_{ij} \geq 0$ and $a_{00}a_{11} \geq 1$.
\end{pro}
The prescription (\ref{Lambda}) for $\Lambda$  is so simple that it seems that for $n > 2$ the corresponding additional conditions for $a_{ij}$ are easy to find. Surprisingly, it is not the case and starting with $n=3$ the general problem is open. We stress that there is an essential difference between $n=2$ and $n>2$. For $n=2$ all positive maps are decomposable. It is no longer true for $n>2$. And there are well known examples of indecomposable maps belonging to a general family (\ref{Lambda}).

Let us recall that a map $\Lambda$ is positive iff for all rank-1 projectors $P$ and $Q$
\begin{equation}\label{DEF}
  {\rm tr}[P \Lambda(Q)] \geq 0\ .
\end{equation}
Taking $P=|x\>\<x|$ and $Q=|y\>\<y|$ one has $ \< x | \Lambda(|y\>\<y|)|x\> \geq 0$ for all $x,y \in \mathbb{C}^n$. Using this definition one may prove the following

\begin{Theorem}[\cite{OSID}]\label{TH1} A map $\Lambda$ defined in (\ref{Lambda}) is positive if and only if $a_{ij} \geq 0$ and for all vectors $x \in \mathbb{C}^n$
\begin{equation}\label{IN}
  \sum_{i=0}^{n-1} \frac{|x_i|^2}{B_i(x)} \leq 1\ ,
\end{equation}
where
\begin{equation}\label{}
 B_i(x) = |x_i|^2 + \sum_{j=0}^{n-1} a_{ij} |x_j|^2\ .
\end{equation}
Moreover, $\Lambda$ is completely positive  if and only if the matrix $D=[d_{ij}]$ such that $d_{ij} = -1$ for $i \neq j$ and $d_{ii} = a_{ii}$ is positive semi-definite.
\end{Theorem}
We stress that an inequality (\ref{IN}) does not provide a solution to our problem. It is just a reformulation of the original definition of positivity for the special class of maps! One may easily check that for $n=2$ an inequality (\ref{IN}) reproduces condition (\ref{Q2}). However, for $n > 2$ we do not know how to translate the  above inequality  into the closed set of conditions upon the matrix elements $a_{ij}$.

\section{Circulant matrices}

Consider now a special case when $a_{ij}$ defines a circulant matrix, i.e. $a_{ij} = \alpha_{i-j}$ (mod $n$). Actually, many well known examples of positive maps belongs to such class (e.g. reduction map, Choi map and its generalizations). We assume that $\alpha_k \geq 0$ for $k=0,\ldots,n-1$ and we denote the corresponding map by $\Lambda[\alpha_0,\ldots,\alpha_{n-1}]$.

\begin{ex}\label{n=2} For $n=2$ denoting  $a_{00}=a_{11}= \alpha_0=:a$ and $a_{01}=a_{10}=\alpha_1 =: b$ formula (\ref{Q2}) reduces to
\begin{equation}\label{}
  a+b \geq 1\ .
\end{equation}
Recall, that $a=0$ and $b=1$ corresponds to the reduction map $R_2(X) = \mathbb{I}_2 {\rm tr}\, X - X$.
\end{ex}

For a circulant matrix Theorem \ref{TH1} reduces to the following

\begin{pro}\label{TH2} A map $\Lambda[\alpha_0,\ldots,\alpha_{n-1}]$ defined in (\ref{Lambda}) is positive if and only if  for all vectors $x \in \mathbb{C}^n$
\begin{equation}\label{CIRC}
  \sum_{i=0}^{n-1} \frac{|x_i|^2}{(\alpha_0 + 1)|x_i|^2 + \sum_{k=1}^{n-1} \alpha_k |x_{i+k}|^2 } \leq 1\ .
\end{equation}
Moreover, $\Lambda$ is completely positive if and only if $\alpha_0 \geq n-1$.
\end{pro}
An inequality (\ref{CIRC}) is known as {\em circulant inequlity} \cite{Yamagami}. In particular taking $|x_0| = \ldots = |x_{n-1}|$ one finds the following necessary condition for positivity of $\Lambda$
\begin{equation}\label{n-1}
  \alpha_0 + \alpha_1 + \ldots + \alpha_{n-1} \geq n-1\ .
\end{equation}
Note, that the above  condition is necessary but not sufficient. Actually, it is sufficient only for $n=2$ (see Example \ref{n=2}).
For $n=3$ a full class of parameters $\alpha_0=a$, $\alpha_1=b$ and $\alpha_2=c$ satisfying circulant inequality (\ref{CIRC}) was derived in \cite{Cho-abc}.

\begin{Theorem}(\cite{Cho-abc})  \label{TH-korea} For $n=3$ a map $\Lambda[a,b,c]$ is positive  if and only if
\begin{enumerate}
\item $ a+b+c \geq 2\ $,
\item if $a \leq 1\ $, then $ \ bc \geq (1-a)^2$.
\end{enumerate}
Moreover, being a positive map  it is indecomposable if and only if
\begin{equation}\label{ind}
  4bc < (2-a)^2\ .
\end{equation}
$\Lambda$ is completely positive if and only if $a\geq 2$.
\end{Theorem}
Hence, for $n=3$ a necessary condition $a+b+c\geq 2$ is supplemented by an extra condition {\em 3}.

\begin{cor} If $a > 1$, then  condition (\ref{n-1}) is necessary and sufficient for positivity of $\Lambda[a,b,c]$.
\end{cor}

\begin{Remark} For $n>3$ a full  set of necessary and sufficient conditions for positivity of $\Lambda[\alpha_0,\ldots,\alpha_{n-1}]$ is not known.
\end{Remark}

\section{Kossakowski construction}

Let us define a set of Hermitian diagonal traceless matrices
\begin{equation}\label{F}
    F_\ell  = \frac{1}{\sqrt{\ell(\ell +1)}} \Big( \sum_{k=0}^{\ell-1} E_{kk} - \ell E_{\ell\ell} \Big) \ ,\ \ \ \ \ell = 1,\ldots,n-1\ .
\end{equation}
These matrices span the Cartan subalgebra of $su(n-1)$. Moreover, ${\rm tr}(F_\alpha F_\beta) = \delta_{\alpha\beta}$.
Define a real $n \times n$ matrix
\begin{equation}\label{aR}
    a_{ij} := \frac{n-1}{n} + \sum_{\alpha,\beta=1}^{n-1} \< e_i|F_\alpha|e_i\> R_{\alpha\beta} \<e_j|F_\beta|e_j\> \ ,
\end{equation}
where $R_{\alpha\beta}$ is an  $(n-1)\times (n-1)$ orthogonal matrix.
Consider now a linear map $\Lambda$ defined by (\ref{Lambda}) with
$a_{ij}$  defined by (\ref{aR}).

\begin{Theorem}[\cite{Kossak-kule}] \label{THK} For any orthogonal matrix $R_{\alpha\beta}$ a linear map $\Lambda$ is positive.
\end{Theorem}

\begin{Remark}
Actually Kossakowski provided more general construction \cite{Kossak-kule}. However, in this paper we restrict our analysis to the special class of diagonal type maps corresponding to (\ref{aR}).
\end{Remark}

Due to the fact that $F_\alpha$ is traceless
for $\alpha=1,\ldots,n-1$, one finds
\begin{equation}\label{doubly}
    \sum_{i=1}^{n-1} a_{ij} = \sum_{j=1}^{n-1} a_{ij} = n-1\ .
\end{equation}
Moreover, since matrix elements $a_{ij} \geq  0$ (it follows from Theorem \ref{THK}) one finds that
\begin{equation}\label{}
    \widetilde{a}_{ij} := \frac{1}{n-1}\, a_{ij}\ ,
\end{equation}
defines a doubly stochastic matrix.

\begin{Remark} A map $\widetilde{\Lambda} := \frac{1}{n-1}\, \Lambda\,$ is unital trace preserving.
\end{Remark}

Consider now an inverse problem: suppose we are given a $n \times n$ matrix $[a_{ij}]$ such that $[\widetilde{a}_{ij}]$ is doubly stochastic. How to check whether $a_{ij}$ is defined {\em via} (\ref{aR})? The answer is given by the following

\begin{pro}[\cite{OSID}] A matrix $[a_{ij}]$ can be represented by (\ref{aR}) if and only if
\begin{equation}\label{osid-a}
    \sum_{k=0}^{n-1} a_{ik} a_{jk} = \delta_{ij} + n-2\ ,
\end{equation}
for $\, i,j=0,\ldots,n-1$.
\end{pro}
Define
\begin{equation}\label{}
  b_{ij} := a_{ij} -1 \ ,
\end{equation}
that is,
\begin{equation}\label{bR}
    b_{ij} =  \sum_{\alpha,\beta=1}^{n-1} \< e_i|F_\alpha|e_i\> R_{\alpha\beta} \<e_j|F_\beta|e_j\>  - \frac{1}{n}\ ,
\end{equation}
One easily proves
\begin{pro} \label{PK1} A matrix $[a_{ij}]$ satisfies (\ref{osid-a}) if and only if matrix $[b_{ij}]$ satisfies
\begin{equation}\label{osid-a}
    \sum_{k=0}^{n-1} b_{ik} b_{jk} = \delta_{ij} \ ,
\end{equation}
for $\, i,j=0,\ldots,n-1$, i.e. $[b_{ij}]$ is an orthogonal matrix.
\end{pro}
Note, that if $[b_{ij}]$ defines an orthogonal matrix, then $|b_{ij}|\leq 1$ and hence $a_{ij} = b_{ij} + 1 \geq 0$.

\begin{cor} A map $\Lambda$ defined in (\ref{Lambda}) is positive if the corresponding $b_{ij}$ defines $n \times n$ orthogonal matrix such that
\begin{equation}\label{doubly-b}
    \sum_{i=0}^{n-1} b_{ij} = \sum_{j=0}^{n-1} b_{ij} = -1\ .
\end{equation}
\end{cor}
It is clear that formula (\ref{bR}) provides an embedding of $O(n-1)$ into $O(n)$, i.e. an orthogonal matrix $R_{\alpha\beta}$ from $O(n-1)$ is mapped into an orthogonal matrix $b_{ij}$ from $O(n)$.

Now, we provide a geometric interpretation of Kossakowski construction. Let $\{ \mathbf{b}^{(0)},\ldots,\mathbf{b}^{(n-1)} \}$ be an orthonormal basis  in $\mathbb{R}^n$ such that
\begin{equation}\label{be}
  (\mathbf{b}^{(i)},\mathbf{e}) = - \frac{1}{\sqrt{n}}\ ,
\end{equation}
where $(\mathbf{a},\mathbf{b})$ denotes the canonical inner product in $\mathbb{R}^n$ and
\begin{equation}\label{e}
\mathbf{e} = \frac{1}{\sqrt{n}}(1,\ldots,1)\ .
\end{equation}
Let us define
\begin{equation}\label{}
  b_{ij} := \mathbf{b}^{(i)}_j \ .
\end{equation}
Clearly, $[b_{ij}]$ defines an orthogonal matrix. Moreover, (\ref{be}) guarantees (\ref{doubly-b}).

\begin{cor} Any Kossakowski map is uniquely defined by an arbitrary orthonormal basis $\{ \mathbf{b}^{(0)},\ldots,\mathbf{b}^{(n-1)} \}$ satisfying (\ref{be}).
\end{cor}

\begin{cor} If $[b_{ij}]$ defines a Kossakowski map, then $[b_{i\pi(j)}]$ defines another Kossakowski map for an arbitrary permutation $\pi \in S_n$.
\end{cor}

Let $\Sigma_\mathbf{e}$ denote an $(n-1)$-dimensional hyperplane in $\mathbb{R}^n$ orthogonal to vector $\mathbf{e}$.
Let $\{ \mathbf{f}^{(1)},\ldots,\mathbf{f}^{(n-1)}\}$ be an arbitrary orthonormal basis in $\Sigma_\mathbf{e}$.  An example of such a basis is provided by
\begin{equation}\label{fF}
  \mathbf{f}^{(\alpha)}_i = \<e_i|F_\alpha|e_i\> \ ,
\end{equation}
where $F_\alpha$ are defined in (\ref{F}). Clearly, $\{ \mathbf{f}^{(0)} := \mathbf{e},\mathbf{f}^{(1)},\ldots,\mathbf{f}^{(n-1)}\}$
defines an orthonormal basis in $\mathbb{R}^n$. Consider now an orthogonal operator $\mathbf{R}$ such that its matrix representation in the basis $\{ \mathbf{f}^{(0)},\mathbf{f}^{(1)},\ldots,\mathbf{f}^{(n-1)}\}$ has the following form
\begin{equation}\label{}
  \mathbf{R}_{00} = -1\ , \ \ \mathbf{R}_{0k}=\mathbf{R}_{k0}=0 \ , \ \ \mathbf{R}_{ij} = R_{ij}\ .
\end{equation}
It is clear that $\mathbf{R}$ represents rotation (or pseudo-rotation) around $\mathbf{e}$.

\begin{pro} Let $[b_{ij}]$ be the matrix representation of $\mathbf{R}$ in the canonical basis in $\mathbb{R}^n$. Then $b_{ij}$ satisfy (\ref{doubly-b}).
\end{pro}
Proof: denote by $\{ \mathbf{e}_0,\ldots,\mathbf{e}_{n-1}\}$ the canonical basis and let
\begin{equation}\label{}
  \mathbf{f}^{(i)} = \sum_{j=0}^{n-1} S_{ij} \mathbf{e}_j\ .
\end{equation}
One has
\begin{equation}\label{bSS}
  b = S^{\rm T} \mathbf{R} S \ .
\end{equation}
and hence
\begin{eqnarray}
  \sum_{i=0}^{n-1} b_{ij} &=& \sum_{k,l=0}^{n-1}\sum_{i=0}^{n-1} S_{ki}\mathbf{R}_{kl} S_{lj} = \sum_{i=0}^{n-1} S_{0 i}\mathbf{R}_{00} S_{0j} + \sum_{\alpha,\beta=1}^{n-1} \sum_{i=0 }^{n-1}S_{\alpha i}{R}_{\alpha\beta} S_{j\beta}  = -1\ ,
\end{eqnarray}
due to
\begin{equation}\label{}
  S_{0 i} = \frac{1}{\sqrt{n}}\ , \ \ \ \sum_{i=0}^{n-1} S_{\alpha i} = 0 \ , \ \ i=0,1,\ldots,n-1\ ; \ \ \alpha= 1,\ldots,n-1\ .
\end{equation}
In particular if $\mathbf{f}^{(i)}$ are defined {\em via} (\ref{fF}), then (\ref{bSS}) reproduces (\ref{bR}).

Consider now the following symmetric set of $n$ vectors $\{ \mathbf{g}^{(0)},\ldots,\mathbf{g}^{(n-1)}\}$ in $\Sigma_\mathbf{\mathbf{e}}$ defined by:

\begin{enumerate}

\item 
they have the same length,

\item the angle `$\phi_n$' between arbitrary two vectors is the same.

\end{enumerate}
One proves that
\begin{equation}\label{}
  \cos\phi_n = - \frac{1}{n-1}\ .
\end{equation}

\begin{Remark} Actually, a set of $n$ vectors $\{ \mathbf{g}^{(0)},\ldots,\mathbf{g}^{(n-1)}\}$ in $\mathbb{R}^{n-1}$ satisfying the above conditions is called an equiangular frame \cite{frame}.
\end{Remark}

\begin{pro} Vectors
$$ \mathbf{b}^{(i)} := \mathbf{g}^{(i)} - \frac{1}{\sqrt{n}} \, \mathbf{e}  \ , $$
such that $|\mathbf{g}^{(i)}|^2 = 1 - \frac 1n$, define an orthonormal basis in $\mathbb{R}^n$ and satisfy (\ref{be}).
\end{pro}



Consider now a special case when the matrix $[a_{ij}]$ defined in (\ref{aR}) is circulant. Formula (\ref{doubly}) implies
\begin{equation}\label{}
  \alpha_0 + \ldots + \alpha_{n-1} = n-1\ .
\end{equation}
In this case Proposition \ref{PK1} reduces to
\begin{pro} \label{PK2} A circulant matrix $a_{ij}= \alpha_{i-j}$ satisfies (\ref{osid-a}) if and only if
\begin{equation}\label{osid-alpa}
    \sum_{k=0}^{n-1} \alpha_{i-k} \alpha_{j-k} = \delta_{ij} + n-2 \ ,
\end{equation}
for $\, i,j=0,\ldots,n-1$.
\end{pro}
Introducing
\begin{equation}\label{}
  \beta_i = \alpha_i - 1\ ,
\end{equation}
one finds
\begin{equation}\label{beta=}
  \beta_0 + \ldots + \beta_{n-1} = -1\ ,
\end{equation}
together with
\begin{equation}\label{osid-beta}
    \sum_{k=0}^{n-1} \beta_{i-k} \beta_{j-k} = \delta_{ij}  \ ,
\end{equation}
for $\, i,j=0,\ldots,n-1$. Clearly, $b_{ij} = \beta_{i-j}$ defines a circulant orthogonal matrix satisfying an additional constraint (\ref{beta=}).

\section{Examples}

\begin{ex} For $n=2$ one has $F_1 = \frac{1}{\sqrt{2}}\, \sigma_z$ and $R = \pm 1$, and hence one easily finds
\begin{equation}\label{}
  R=1 \ \rightarrow \ [a_{ij}] = \mathbb{I}_2 \ ; \ \   R=-1 \ \rightarrow \ [a_{ij}] = \sigma_x \ .
\end{equation}
\end{ex}

\begin{ex}\label{E-n=3} For $n=3$
\begin{equation}\label{}
  F_1 = \frac{1}{\sqrt{2}} \left( \begin{array}{ccc} 1 & 0 & 0 \\ 0 & -1 & 0 \\ 0 & 0 & 0 \end{array} \right) \ ; \ \ \
   F_2 = \frac{1}{\sqrt{6}} \left( \begin{array}{ccc} 1 & 0 & 0 \\ 0 & 1 & 0 \\ 0 & 0 & -2 \end{array} \right) \ ,
\end{equation}
and
\begin{equation}\label{}
  [R_{\alpha\beta}] = \left( \begin{array}{cc} \cos \phi & \sin\phi \\ - \sin \phi & \cos\phi  \end{array} \right) \ .
\end{equation}
Interestingly, in this case one finds that the matrix $[a_{ij}]$ is circulant. Denoting $a:= a_{00}, b:=a_{01}$ and $c:=a_{02}$ one obtains
\begin{eqnarray}\label{abc-phi}
    a &=&  \frac 23 ( 1 + \cos\phi) \ , \nonumber \\
    b &=&  \frac 13 ( 2 - \cos\phi - \sqrt{3} \sin\phi) \ , \\
    c &=&  \frac 13 ( 2 - \cos\phi + \sqrt{3} \sin\phi) \nonumber \ .
\end{eqnarray}
Let us observe that introducing $\widetilde{a} = a-1, \widetilde{b}=b-1$ and $\widetilde{c}=c-1$ the above family of maps is uniquely characterized by a circulant orthogonal matrix
\begin{equation}\label{}
  [b_{ij}] =  \left( \begin{array}{ccc} \widetilde{a} & \widetilde{b} & \widetilde{c} \\   \widetilde{c} & \widetilde{a} & \widetilde{b} \\ \widetilde{b} & \widetilde{c} & \widetilde{a} \end{array} \right) \ ,
\end{equation}
with $\widetilde{a} + \widetilde{b} + \widetilde{c} =-1$. Interestingly, the well known maps: Choi maps $\Lambda[1,1,0]$, $\Lambda[1,0,1]$ and the reduction map $\Lambda[0,1,1]$ have the following representation in terms of the matrix $[b_{ij}]$:
\begin{equation}\label{}
  \left( \begin{array}{ccc} 0 & 0 & -1 \\ -1 & 0 & 0 \\ 0 & -1 & 0 \end{array} \right) \ ; \ \
  \left( \begin{array}{ccc} 0 & -1 & 0 \\ 0 & 0 & -1 \\ -1 & 0 & 0 \end{array} \right)\ ; \ \
  \left( \begin{array}{ccc} -1 & 0 & 0 \\ 0 & -1 & 0 \\ 0 & 0 & -1 \end{array} \right)\ ,
\end{equation}
that is, up to a sign they correspond to circulant permutation matrices.
\end{ex}

\begin{Remark} For $n=2$ and $n=3$ all Kossakowski maps are characterized by a circulant matrix $[a_{ij}]$. It is no longre true for $n > 3$.
\end{Remark}

\begin{Remark} Let us observe that parameters $a,b,c$ defined in  (\ref{abc-phi}) are compatible with Theorem \ref{TH-korea}.
Note, that maps defined via (\ref{abc-phi}) belong to the boundary of a set of positive maps defined by two equalities in conditions {\em 1}. and {\em 2}. of   Theorem \ref{TH-korea}, that is,
\begin{equation}\label{}
  a+b+c=2\ ; \ \ \ bc = (1-a)^2\ .
\end{equation}
 Detailed analysis of the structure of these maps was performed in \cite{FilipI}.
\end{Remark}


\begin{ex}  \label{E-n=4} For $\, n= 4$ one has the following circulant orthogonal $[b_{ij}]$ matrix:
$\widetilde{a} = b_{00}$, $\widetilde{b} = b_{01}$, $\widetilde{c} = b_{02}$ and $\widetilde{d}=b_{03}$
satisfying
\begin{equation}\label{4-0}
    \widetilde{a} + \widetilde{b} + \widetilde{c} + \widetilde{d} =-1\ .
\end{equation}
Orthogonality conditions imply
\begin{equation}\label{3-1}
    \widetilde{a}^2 + \widetilde{b}^2 + \widetilde{c}^2 + \widetilde{d}^2 = 1 \ , \ \ \ \widetilde{a}\widetilde{c} + \widetilde{b}\widetilde{d} = 0  , \ \ \ (\widetilde{a} + \widetilde{c})(\widetilde{b} + \widetilde{d}) = 0 \ .
\end{equation}
Therefore, we have two classes of
admissible parameters $\{\widetilde{a} , \widetilde{b} , \widetilde{c} , \widetilde{d}\}$ constrained by
\begin{equation}\label{4I}
     \widetilde{a} + \widetilde{b} + \widetilde{c} + \widetilde{d} =-1 \ , \ \  \widetilde{a}^2 + \widetilde{b}^2 + \widetilde{c}^2 + \widetilde{d}^2 = 1  \ , \ \  \widetilde{b} + \widetilde{d} = 0  \ ,
\end{equation}
and
\begin{equation}\label{4II}
     \widetilde{a} + \widetilde{b} + \widetilde{c} + \widetilde{d} =-1 \ , \ \  \widetilde{a}^2 + \widetilde{b}^2 + \widetilde{c}^2 + \widetilde{d}^2 = 1  \ , \ \  \widetilde{a} + \widetilde{c} = 0  \ ,
\end{equation}
Equivalently, the above conditions may be rewritten as follows
\begin{equation}\label{4Ia}
   \widetilde{a}^2 + \widetilde{b}^2 + \widetilde{c}^2 + \widetilde{d}^2 = 1  \ , \ \   \widetilde{a} + \widetilde{c} = -1 \ , \ \ \widetilde{b} + \widetilde{d} = 0  \ ,
\end{equation}
and
\begin{equation}\label{4IIa}
\widetilde{a}^2 + \widetilde{b}^2 + \widetilde{c}^2 + \widetilde{d}^2 = 1  \ , \ \  \widetilde{a} + \widetilde{c} = 0  \ , \ \  \widetilde{b} + \widetilde{d} = -1\ .
\end{equation}
They describe two circles: the intersection of 3D sphere with two planes. Again, characteristic well known maps $\Lambda[1,1,1,0], \Lambda[1,1,0,1], \Lambda[1,0,1,1]$ and $\Lambda[0,1,1,1]$  (up to a sign) correspond to circulant permutation matrices:
\begin{equation*}\label{}
  \left( \begin{array}{cccc} 0 & 0 & 0& -1 \\ -1 & 0 & 0 &0  \\ 0 & -1 & 0 & 0 \\ 0 & 0 & -1 & 0  \end{array} \right) \ ,
  \left( \begin{array}{cccc} 0 & 0 & -1& 0 \\ 0 & 0 & 0 &-1  \\ -1 & 0 & 0 & 0 \\ 0 & -1 & 0 & 0  \end{array} \right) \ ,
  \left( \begin{array}{cccc} 0 & -1 & 0& 0 \\ 0 & 0 & -1 &0  \\ 0 & 0 & 0 & -1 \\ -1 & 0 & 0 & 0  \end{array} \right) \ ,
  \left( \begin{array}{cccc} -1 & 0 & 0& 0 \\ 0 & -1 & 0 &0  \\ 0 & 0 & -1 & 0 \\ 0 & 0 & 0 & -1  \end{array} \right)\ .
\end{equation*}

\end{ex}

\begin{ex} For $n=5$ one has the following circulant orthogonal $[b_{ij}]$ matrix:
$\widetilde{a} = b_{00}$, $\widetilde{b} = b_{01}$, $\widetilde{c} = b_{02}, \widetilde{d}=b_{03}$ and $\widetilde{e} = b_{04}$
satisfying
\begin{equation}\label{5-0}
    \widetilde{a} + \widetilde{b} + \widetilde{c} + \widetilde{d} + \widetilde{e}=-1\ .
\end{equation}
Orthogonality conditions imply
\begin{equation}\label{5-1}
    \widetilde{a}^2 + \widetilde{b}^2 + \widetilde{c}^2 + \widetilde{d}^2 + \widetilde{e}^2 = 1\ , \ \
    \widetilde{a}\,\widetilde{e} + \widetilde{b}\,\widetilde{a} + \widetilde{c}\,\widetilde{b} + \widetilde{d}\,\widetilde{c} +     \widetilde{e}\,\widetilde{d} = 0\ .
\end{equation}
One easily checks that the remaining orthogonality conditions are not independent from (\ref{5-0}) and (\ref{5-1}). The corresponding set of admissible parameters $\{ \widetilde{a} , \widetilde{b} , \widetilde{c} , \widetilde{d} , \widetilde{e}\}$ is 2-dimensional but its shape is not very transparent  (\ref{5-0}) and (\ref{5-1}).
\end{ex}

\section{Circulant case --- a complementary parametrization}

Let us recall that if $a_{ij} = \alpha_{i-j}$  defines a circulant matrix, then its eigenvalues are given by
\begin{equation}\label{DFT}
    \lambda_{k} = \sum_{l=0}^{n-1} \omega^{-kl} \alpha_l\ , \ \ \ k=0,\ldots, n-1\ ,
\end{equation}
and the corresponding eigenvectors read
\begin{equation}\label{}
  \mathbf{x}_k = (1,\omega^k,\omega^{2k},\ldots,\omega^{(n-1)k})^{\rm T}\ ,
\end{equation}
where $\omega= e^{2\pi i/n}$. Two sets $\{\alpha_0,\ldots,\alpha_{n-1}\}$ and $\{\lambda_0,\ldots,\lambda_{n-1}\}$ are related by the discrete Fourier transform.
Note that
\begin{equation}\label{z0}
  \lambda_0 = \alpha_0 + \ldots + \alpha_{n-1} = n-1\ .
\end{equation}
Consider now a circulant orthogonal matrix $b_{ij} = \beta_{i-j}$ with $\beta_k = \alpha_k -1$. The corresponding eigenvalues $\mu_k$ of $[b_{ij}]$ are defined by
\begin{equation}\label{}
  \mu_0 = \lambda_0 - n = -1\ , \ \ \ \mu_\alpha = \lambda_\alpha\ , \ \ \alpha=1,\ldots,n-1\ .
\end{equation}
Now, since $[b_{ij}]$ is orthogonal one has $|\mu_k|=1$ and hence

\begin{pro} Real parameters $\{\alpha_0,\ldots,\alpha_{n-1}\}$ satisfy (\ref{osid-alpa}) if and only if $|\lambda_\alpha|=1$ for $\alpha = 1,\ldots,n-1$.
\end{pro}

This way we obtain a new parametrization of a set of admissible circulant matrices $[a_{ij}]$ by phases of $\lambda_\alpha = e^{i\phi_\alpha}$. Due to $\lambda_k = \lambda_{n-k}^*$ one has two cases:

\begin{enumerate}

\item if $n=2m+1$, then we have $m$ independent phases $\lambda_1 = e^{i\phi_1}, \ldots , \lambda_m = e^{i\phi_m}$.

\item if $n=2m+2$, then we have $m$ independent phases $\lambda_1 = e^{i\phi_1}, \ldots , \lambda_m = e^{i\phi_m}$ and one real parameter $\lambda_{m+1} = \pm 1$.

\end{enumerate}

\begin{ex} For $n=3$ putting $\lambda_1 = e^{i\phi}= \lambda_{2}^*$ one finds
\begin{eqnarray}\label{}
    a &=& \frac 13 ( 2 +  \lambda_1 + \lambda_1^*) = \frac 23 ( 1 + \cos\phi) \ , \nonumber \\
    b &=& \frac 13 ( 2 + \omega \lambda_1 + \omega^* \lambda_1^*) = \frac 13 ( 2 - \cos\phi - \sqrt{3} \sin\phi) \ , \\
    c &=& \frac 13 ( 2 + \omega^* \lambda_1 + \omega \lambda_1^*) = \frac 13 ( 2 - \cos\phi + \sqrt{3} \sin\phi) \nonumber \ ,
\end{eqnarray}
due to $\omega =e^{2\pi i/3} = \frac 12 (-1 + i \sqrt{3})$. This reproduces result of Example \ref{E-n=3}.
\end{ex}

\begin{ex} For $n=4$ if $\lambda_1 = e^{i\phi}= \lambda_3^*$ and $\lambda_2 = 1$ one finds
\begin{equation}\label{}
 a = \frac 12(2 + \cos\phi) \ , \ \  b = \frac 12(1 - \sin\phi) \ , \ \ c = \frac 12(2 - \cos\phi) \ , \ \ d = \frac 12 (1 + \sin\phi) \ ,
\end{equation}
and similarly if $\lambda_1 = e^{i\psi}= \lambda_3^*$ and $\lambda_2=-1$ one has
\begin{equation}\label{}
    a = \frac 12(1 + \cos\psi) \ , \ \  b = \frac 12(2 - \sin\psi) \ , \ \ c = \frac 12(1 - \cos\psi) \ , \ \ d = \frac 12 (2 + \sin\psi) \ .
\end{equation}
Note, that for $\lambda_2=1$ one has $b+d=1$, whereas for $\lambda_2=-1$ one has $b+d=2$. This way we reproduced two classes from Example \ref{E-n=4}.
\end{ex}

\begin{cor} It is therefore clear that

\begin{enumerate}

\item if $n=2m+1$, then a set of admissible parameters defines $m$-dimensional torus $\mathbb{T}_m$. Note that $O(n-1) = O(2m)$ and a single torus $\mathbb{T}_m$ corresponds to a maximal commutative subgroup of $SO(2m)$.

\item if $n=2m+2$, we have two $m$-dimensional tori $\mathbb{T}_m$ and $\mathbb{T}_m'$. Torus $\mathbb{T}_m$ corresponds to a maximal commutative subgroup of $SO(2m+1)$ whereas $\mathbb{T}_m'$ is defined by composing $\mathbb{T}_m$  with reflection, that is, $g \in \mathbb{T}_m'$ iff $-g \in \mathbb{T}_m$ (cf. \cite{JPA-Bell}).

\end{enumerate}

\end{cor}

\begin{cor} Positive maps $\Lambda[\alpha_0,\ldots,\alpha_{n-1}]$ are invertible.  It follows from the fact that
\begin{equation}\label{}
  |{\rm det}[a_{ij}]| = |\lambda_0 \ldots \lambda_{n-1}  | = n-1 \neq 0\ .
\end{equation}
Note, however, that the inverse $\Lambda^{-1}[\alpha_0,\ldots,\alpha_{n-1}]$ is no longer positive.
\end{cor}

\section{Conclusions}

We analyzed a class of  positive maps introduced by Kossakowski \cite{Kossak-kule}. It turns out that these maps display interesting geometric features. In particular its maximal commutative subset --- $\Lambda[\alpha_0,\ldots,\alpha_{n-1}]$ --- corresponding to circulant matrices $[a_{ij}]$ is parameterized by tori which defines maximal commutative subgroups of the orthogonal group. For further properties of these maps like (in)decomposability and/or optimality see also \cite{JPA-Bell}. It is clear that {\em via} Choi-Jamio{\l}kowski isomorphism can one provide a similar analysis in terms of entanglement witnesses (see  \cite{TOPICAL}  for the recent review).

\section*{Acknowledgements}

This paper was partially supported by the National Science Center project DEC-
2011/03/B/ST2/00136. I thank Andrzej Kossakowski  for the  fruitful and inspiring
discussions and Andrzej Jamio{\l}kowski for pointing out Ref. \cite{frame}.

\end{document}